\documentstyle[12pt,aaspp4,tighten]{article}

\def\etal{{\it et al.~}}

\begin{document}

\title{On the time variation of $c$, $G$, and $h$ and the dynamics of the cosmic expansion}

\author{Ari Buchalter\altaffilmark{1}}
\altaffiltext{1}{aribuch2@aol.com}

\begin{abstract}

Several authors have recently explored the idea that physical constants
such as $c$ and $G$ might vary over time and have formulated theories
describing this variation that can address a range of cosmological 
problems. Such work typically invokes a generic parameterization which assumes
a power-law variation with the expansion scale factor, $R$.
This work offers alternative, physically motivated definitions
for the parameters $c$, $G$, and $h$ 
based on the Machian premise that these dimensional quantities
reflect global dynamics of the expansion geometry.
Together with a postulated conservation law and equations of motion,
the implications of this theory for Friedmann models are examined, and found
to yield several interesting conclusions including:
(1) natural solutions to the horizon, flatness, and lambda problems,
(2) the prediction of a flat, $\Omega_0 = 1$ universe,
(3) different forms for some cosmological scaling laws, 
(4) an apparent fit to
observations of Type Ia supernovae without invoking a cosmological constant, 
(5) equivalence between our Universe and a black hole 
and apparent consistency of the model with the Holographic Principle, and (6) 
potentially testable predictions 
for the time variation of physical parameters, 
including values for $\dot{c_0}$ and $\dot{h_0}$ that are small but non-zero today
and a value for $\dot{G}$ that was negative and nonzero during radiation domination 
and decayed to effectively zero upon the epoch of matter domination.
While this work does not attempt
to provide the complete theoretical foundation that must ultimately underlie
any theory that could naturally marry traditional physics with the notion
of time-varying physical parameters, it is written in the hope that it might
stimulate further progress towards this end.

\end{abstract}

\section{Introduction}

For over three centuries, theories of physics have been characterized by
the existence of physical constants.  Despite their ubiquity, however,
the significance of these dimensional parameters
remains relatively unexplored. While some, such as the Rydberg,
have historically been revealed to be comprised
of other more fundamental quantities, others such as the gravitational constant
$G$, the speed of light $c$, and Planck's constant $h$, are of seemingly
primary significance. If precedence suggests that constants
might be viewed as placeholders for as yet undiscovered
physics in our theories, then one of the challenges of physics is surely to reveal
the physical meaning of these parameters. Indeed, it might be generally supposed 
that as physics progresses,
fewer constants will be required as more fundamental theories are put forth--that
a true ``theory of everything'' might contain no such quantities, explaining nature 
from first principles alone. In practice, of course,
the Universe sometimes reveals complexities faster than
theorists can explain them, and new constants are born. One such example is the 
cosmological constant $\Lambda$, introduced (and later bemoaned) by Einstein 
in an attempt to arrive at a steady-state characterization of the 
Universe\footnote{A $\Lambda$-dominated model was later shown to be, in fact, unstable.}, 
and which in light of recent observations is now nearly universally accepted (Reiss \etal 1998;
Perlmutter \etal 1999; Knop \etal 2003; Bennett \etal 2003; Tegmark \etal 2003).

Several papers in recent years have explored the significance of physical ``constants''
(henceforth termed physical parameters)
by assuming an ad hoc form for their possible variation over time and 
investigating the corresponding 
implications. For example, Albrecht \& Magueijo (1999) and Barrow (1999) show
that simple varying speed of light (VSL) theories offer an alternative to inflation
for solving several of the problems associated with the Standard Big Bang (SBB) cosmology, 
such as the horizon, flatness, and lambda problems. Barrow (2003)
demonstrates, however, that these simple models may face challenges, for example, in solving the
isotropy problem and generating constant-curvature fluctuations, 
clearly highlighting the need for additional work to make such models viable.

The present work seeks to explore the significance of physical parameters from the Machian
standpoint that these dimensional quantities reflect global characteristics of the Universe.
From this premise, we will formulate two simple postulates relating $c$ and $G$
to the dynamics of the expansion geometry, offering a physical motivation for the
meaning of these parameters.\footnote{If physical parameters reflect global properties
of the Universe, it is only natural that they {\em should} vary in time in accordance
with cosmological dynamics, i.e., the expansion. 
As there are obviously no physical constants in mathematics, one may suppose that it
dimensionless, rather than dimensional, quantities that are in fact invariant.
Indeed, since these assumptions will be shown
to necessitate the existence of a preferred cosmological frame, 
it may be 
philosophically more reasonable to suppose that the distance traveled by a photon per unit time
in such a frame frame might vary, than that the number 3 should evolve
into 1 or 59$\sqrt{\pi}$. As a dimensionless quantity, the fine-structure parameter $\alpha$
is assumed to be constant in this model.} A definition for $h$, which will also be linked to the
expansion geometry, will follow from these postulates.
Together with two additional postulates
(a conservation law and equations of motion), we begin construction of a theory 
which appears to have several interesting features. A limited subset of these is presented here,
including: (1) natural solutions to the horizon, flatness, lambda, and
potentially other SBB problems, (2) the prediction of a flat, $\Omega_0 = 1$ universe,
(3) different forms for some cosmological scaling laws, 
(4) an apparent
fit to observations of Type Ia supernovae without invoking a cosmological constant, 
(5) equivalence between our Universe and a black hole 
and apparent consistency of the model with the Holographic Principle, and (6) 
potentially testable predictions 
for the time variation of physical parameters, 
including values for $\dot{c_0}$ and $\dot{h_0}$ that are small but non-zero today
and a value for $\dot{G}$ that was negative and nonzero during radiation domination 
and decayed to effectively zero upon the epoch of matter domination.
Note that in the context of this theory, any predicted
time variation (or lack thereof) in the physical parameters arises 
naturally from their physical definitions, rather than from an assumed parameterization.

Section \ref{section1} of this paper puts forth the postulates for the
physical definitions of $c(t)$ and $G(t)$ and two additional postulates: 
a conservation equation and equations of motion.
Section \ref{section2} explores some of the resulting cosmological implications in Friedmann
models, including specification of $G(t)$, solutions to SBB problems, 
specification of the expansion geometry,
modified scaling laws, calculation of the luminosity distance, 
and application to Type Ia SN observations.
Section \ref{section3} investigates the quantitative behavior of $c(t)$ and
$G(t)$ and draws the mathematical connections between our Universe and a black hole.
Section \ref{section4} puts forth a corollary for the physical definition of $h(t)$ and
further develops the notion of ``black hole equivalence'' via thermodynamic analogy
and apparent consistency with the Holographic Principle. Additional thoughts and
a discussion on the resolution of this model with Special and General Relativity 
are presented in the Conclusion.

It is clear that any theory which motivates the existence of physical parameters
and/or introduces their time variation 
must contend not only with the classical SBB problems and with
the strong observational limits that have been measured
(Rich 2003, Magueijo 2003), but ultimately must produce a compelling model 
of physics that expands our 
understanding of natural phenomena, rather than simply interesting 
heuristic exercises. Like most advances, however, the development of such a theory is 
likely to occur in small steps. The present exercise is merely an attempt to
take one of those steps in the hopes that it might stimulate further
discussion. It will focus on deriving a set of conclusions based on
a series of postulates, but does not attempt to provide a complete theory nor to  
explore the entire range of implications.

\section{Machian Interpretations of $c(t)$ and $G(t)$} \label{section1}

\subsection{The Speed of Light}

Numerous VSL theories have been put forth in the past century, most recently by authors
who note that such theories pose an alternative to inflation for addressing
some of the problems with SBB cosmology
(e.g., Moffat 1993, 2002; Albrecht \& Magueijo 1999; 
Barrow 1999; see Magueijo 2003 for a review).
Typically, these authors have chosen to model the time variation of $c$ as
\begin{equation}
c(t) \propto {R^n}
\label{powerc}
\end{equation}
where $R=R(t)$ is the expansion scale factor in units of length and $n$ is a constant.
Assuming the Friedmann equations retain the same form with $c(t)$ varying
Albrecht \& Magueijo (1999) and Barrow (1999)
show that the flatness and horizon problems can be solved if $n \leq -1$
in a radiation-dominated universe or $n \leq -1/2$ in a matter-dominated
universe, and further demonstrate potential solutions to the lambda,
homogeneity, and isotropy problems.

The functional form of equation (\ref{powerc}), however, provides little insight into
{\em why} $c$ is varying. If we adopt a Machian stance and assert that physical
parameters such as $c$, $G$, and $h$ reflect global characteristics of the Universe, 
we may suppose $c$ to be fundamentally tied to the cosmic expansion.
The simplest assumption consistent with this assertion would be to take

{\bf Postulate~1.}
\begin{equation}
c = c(t) \equiv \dot{R}(t) 
\end{equation}
i.e., defining $c(t)$ as the rate at which the scale factor is changing.
Before turning to a discussion of the consequences and implications
of this postulate, we lay out some additional postulates of the present
theory and reserve their collective discussion for the sections that follow.

\subsection{The Gravitational Constant}

$G$ holds the distinction of being the longest-standing constant in physics.
Recent work by Barrow (1999), Barrow \& Magueijo (1999), and 
Barrow, Magueijo \& Sandvik (2002)
has built upon the theory of Brans-Dicke (1961) and Bekenstein (1982)
to formulate extensions of general relativity that incorporate variation
in $G$ and describe the behavior of Friedmann universes in such a model.

Barrow (1999) shows that time variation in $c$ and $G$ can solve the 
lambda problem by assuming $c(t)$ follows
equation (\ref{powerc}) and a similar form for $G$,
\begin{equation}
G(t) \propto R^q
\label{powerg}
\end{equation}
where $q$ is a constant.

However, if we again invoke the Machian 
notion that $G$ describes some fundamental characteristic of the Universe, we 
can arrive at a more physically motivated prescription. On purely dimensional grounds, 
$G$ can be interpreted as 
the second time derivative of a volume per unit mass. 
Thus, we assume

{\bf Postulate~2.}
\begin{equation}
G = G(t) \equiv \mu \frac{d^2 (V/M) }{d t^2} 
\label{G}
\end{equation}
where $M$ is the mass of the universe, $\mu$ is a constant to be determined below and
$V$ is the proper volume of the observable Universe, which we may write as
\begin{equation}
V = \nu {d_H}^3
\end{equation}
where $d_H$ is the proper distance to the particle horizon and $\nu$ is a geometric
factor determined by the curvature (e.g., $\nu = 4\pi/3$ for $k=0$, or $2\pi ^2$
for $k=1$).

Postulates 1 and 2 would seem to imply that massless particle
propagation and gravity arise
as a consequence of the expansion dynamics, which in turn depend on the curvature and
the energy constituents of the Universe. Before interpreting Postulates 1 and 2 in the 
context of Friedmann models, we put forth two additional postulates:

{\bf Postulate~3.}
\begin{equation}
\frac{dM}{dt} = 0. 
\label{dM}
\end{equation}

{\bf Postulate~4.}
\begin{equation}
{\rm The~Friedmann~equations~retain~the~same~form~under~postulates~1-3.}  
\label{P4}
\end{equation}

Postulate 3, which states that the total mass of the Universe is constant, 
bears a deep significance with respect to the current model, and we will
later explore its explicit connection to postulates 1 and 2 and some of the implications for
thermodynamic relationships. Postulate 4 was put forth by Albrecht \& Magueijo (1999)
and proven in the context of a specific VSL theory. They noted that a varying $c$
does not introduce changes to curvature and therefore should not affect
the Einstein equations. It does lead, however,
to a non-covariant theory and thus requires
the selection of a preferred time coordinate, chosen in the present model to be the comoving
proper time in the cosmological frame. 
A more general proof of postulate 4 incorporating $c$ and $G$ may well be possible via
an appropriate generalization of General Relativity,
but is beyond the current scope. However,
it is noted that Moffat (2003) derives a general bimetric gravity theory wherein both
$G$ and $c$ vary, and which produces some of the same qualitative features of the
present model, though differing in detail.
It should also be noted that Brans-Dicke theory 
is not consistent with Postulates 1, 2, and 4, since it attempts to satisfy the standard
energy conservation equation with $\dot{G} \neq 0$ and thus introduces additional terms
into the Einstein equations. The current model (which, as may be expected, violates
energy conservation due to varying $c$) instead introduces changes to the
conservation equations via postulate 3, which will be shown to be entirely consistent with
Postulates 1 and 2. Note that Postulates 1 and 2 assume no spatial variation in $c(t)$ or
$G(t)$.

\section{FRW Models with $G(t)$ and $c(t)$} \label{section2}

\subsection{Specification of $G(t)$} 

We begin by writing
the Robertson-Walker metric for a spatially
homogeneous and isotropic universe as
\begin{equation}
ds^2 = c^2(t)dt^2 - R^2(t) \left(\frac{dr^2}{1-kr^2} + r^2 d\psi^2 \right)
\label{RW}
\end{equation}
where $t$ is the comoving proper time, $r$ is the dimensionless
comoving coordinate, $k=0,+1,-1$ for zero, positive, or negative curvature, 
$d\psi^2 = d\theta^2 +\sin^2\theta d\phi^2$, and 
the time-variation of $c(t)$ is explicit. Equation (\ref{RW}) can be shown to follow
from the analogous generalization for $c(t)$ in the ordinary Minkowski metric
for flat space.

Using Postulate 1, we can then see that $R(t)$ defines the proper horizon distance:
\begin{equation}
R(t) \equiv \int_0^t {c(t)dt} = R(t) \int_0^{r_H(t)} \frac{dr}{\sqrt{1-kr^2}}  \equiv d_H(t).
\label{horizon}
\end{equation}
In other words, $c=\dot{R}$ implies that at a given time $t$ an observer 
at $r=0$ can just receive light signals emitted 
at $t=0$ from the horizon (at a comoving distance of $r_H (t) =1$).
Note that this result holds generally for any choice of cosmology.

Equation (\ref{horizon}) differs, for
example, from the familiar constant-$c$ result for $d_H$ in a flat universe with 
zero cosmological constant,
\begin{equation}
d_H(t) \propto ct \propto R^{1/m}
\;\;\;\;\;\;\;\;\;\;\;\; ({\rm constant}~c).
\label{SBBdH}
\end{equation}
In the SBB model, a universe consisting of a fluid  with the equation of state
$p=w \rho c^2$
where $p$ is the fluid pressure, $\rho$ is the density, and $w$ is a constant, obeys
$R \propto t^m$ where $m= 2/[3(1+w)]$, so that equation (\ref{SBBdH})
says that for any $w > -1/3$, the horizon grows faster than $R(t)$
(e.g., $d_H \propto R^{3/2}$ for matter domination and 
$d_H \propto R^{2}$ for radiation domination). This leads to the so-called
horizon problem, given the uniformity of the microwave background radiation across
causally disconnected regions in the SBB model. 
The choice of $c=\dot{R}$ which leads to $d_H = R$ just critically solves
the horizon problem, indeed, for any cosmology. Taking $c =\dot{R} \propto t^{m-1}$
means that for any $m<1$, 
$c$ decreases with time, implying a perpetually ``superluminal'' (relative to $c$ today)
expansion history that resolves the horizon problem in a manner analogous to inflation
(in the SBB model, this would correspond to the choice of $w>-1/3$, though we will
later show that the SBB relation between $w$ and $m$ does not hold in the present model).

Having defined $d_H$, we can invoke postulates 2 and 3 to obtain
\begin{equation}
G(t) = \mu \frac{d^2(V/M)}{dt^2} = 
\frac{\mu \nu}{M} \left( 6 R \dot{R}^2 + 3 R^2 \ddot{R} \right).
\label{G2}
\end{equation}
Invoking Postulate 4, we write the Friedmann equations as
\begin{equation}
\frac{\dot{R}^2}{R^2} = \frac{8{\pi}G(t)}{3}\rho - \frac{k \dot{R}^2}{R^2}
\label{Friedmann1}
\end{equation}
\begin{equation}
\frac{\ddot{R}}{R} = \frac{-4{\pi}G(t)}{3} \left(\rho + \frac{3p}{\dot{R}^2}\right)
\label{Friedmann2}
\end{equation}
where a dot denotes a derivative with respect to
comoving proper time, and $\rho$ and $p$ are the
total cosmological density and pressure, which may contain contributions from various
species (matter, radiation, cosmological constant\footnote{We define
the cosmological constant, 
$\lambda$ to be a true dimensionless constant, as opposed to the traditionally
assumed $\Lambda$ which has units of inverse length squared, so that $\lambda \equiv 
\Lambda R^2$. Often, the cosmological constant density, $\rho_\lambda = \lambda c^2 / 8
\pi G R^2$ is treated separately, resulting in the addition of the term
$\lambda c^2 / 3R^2$ to the right-hand side of equations (\ref{Friedmann1}) 
and (\ref{Friedmann2}).\label{foot1}}
$\lambda$, etc.) denoted by the index $i$, such that
\begin{equation}
\rho = \sum_i \rho_i; \;\;\;\;\;\;\;\;\;\;\; 1 = \sum_i \eta_i \;\;\;\;\; {\rm where} 
\;\;\;\;\; \eta_i = \frac{\rho_i}{\rho}. 
\end{equation}
We shall assume the equation of state has the form
\begin{equation}
p = \sum_i p_i = w \rho c^{2}(t) = w \rho \dot{R}^2
\label{eos}
\end{equation}
and define $w = \sum_i w_i \rho_i / \rho = \sum_i w_i \eta_i$.
Note that equation (\ref{Friedmann1}) can be recast in the familiar form
\begin{equation}
1 = \Omega - \Omega_k
\label{Omega}
\end{equation}
where $\Omega = \Omega_{\rm m} + \Omega_{\rm r} + \Omega_{\lambda}$ plus possible terms for 
other species, and
\begin{equation}
\Omega_{\rm m} = \frac{8\pi G(t)}{3} \frac{\rho_{\rm m}R^2}{\dot{R}^2};  \;\;\;\;\;
\Omega_{\rm r} = \frac{8\pi G(t)}{3} \frac{\rho_{\rm r}R^2}{\dot{R}^2};  \;\;\;\;\;
\Omega_k = \frac{k c^2(t)}{\dot{R}^2} = k; \;\;\;\;\;
\Omega_\lambda  =  \frac{\lambda c^2(t)}{3\dot{R}^2} =\frac{\lambda}{3}
\label{OmegaDefs}
\end{equation}
where a subscripted m or r denotes matter or radiation.

Applying our expressions for $c(t)$ and $G(t)$ to 
equation (\ref{Friedmann1}), we have
\begin{equation} 
\frac{\dot{R}^2}{R^2} = \frac{8\pi\mu}{3} \left( \frac{6 R \dot{R}^2 + 3 R^2 \ddot{R}}
{R^3} \right) - \frac{k \dot{R}^2}{R^2} 
\end{equation}
where $\nu$ and $M$ have canceled with the density term, $\rho = M/{\nu R^3}$. This reduces to
\begin{equation} 
1 = {8\pi\mu} \left( 2 + \frac{ R \ddot{R}}{\dot{R}^2} \right) - k.
\label{modFriedmann}
\end{equation}
Applying our expressions for $c(t)$ and $G(t)$ to 
to the second Friedmann equation (\ref{Friedmann2})
and using equation (\ref{eos}) yields
\begin{equation}
\frac{\ddot{R}}{R} = {-4\pi \mu} (1+3w) \left(2\frac{\dot{R}^2}{R^2} + 
\frac{\ddot{R}}{R}\right)
\end{equation}
which can be rewritten as
\begin{equation}
\frac{\ddot{R}R}{\dot{R}^2} = \frac{-8\pi \mu(1+3w)}{1+4\pi \mu (1+3w)} \equiv -q.
\label{q2}
\end{equation}
Substituting equation (\ref{q2}) into equation (\ref{modFriedmann}) and noting
from equations (\ref{Omega}) and (\ref{OmegaDefs}) that $\Omega = 1+k$
allows us to solve for $\mu$
\begin{equation}
\mu = \frac{1}{8\pi}\left( \frac{\Omega}{2  - \Omega (1+3w)/2 }\right)
\label{mu}
\end{equation} 
which equals $1/12\pi$ or $1/8\pi$ for flat universes filled with matter ($w=0$)
or radiation ($w=1/3$), respectively.
Plugging the expression for $\mu$ into equation (\ref{q2}) we have
\begin{equation}
q = - \frac{\ddot{R}R}{\dot{R}^2} = \frac{\Omega}{2}(1+3w).
\label{q}
\end{equation}

\subsection{The Flatness and Horizon Problems}

The results above clearly carry a range of implications for the behavior
of Friedmann universes, some of which we now explore. Earlier, it was demonstrated
that the current model solves the horizon problem (we shall henceforth
denote the current model as the Varying Physical Parameter (VPP) model). 
It is now easy to demonstrate
that the VPP model also addresses the flatness and lambda problems. 
 
For a $\lambda = 0$ model with constant $c$, the SBB model states that
\begin{equation}
\Omega - 1 = \Omega_k = \frac{k c^2}{\dot{R}^2}  \;\;\;\;\;\;\;\;\;\;\;\; ({\rm constant}~c).
\end{equation}
For any SSB equation of state with $w>-1/3$, $\dot{R}$ 
is a decreasing function of time. Hence, the curvature term
tends to zero for small $t$, requiring that $\Omega$ be extremely close to 1 in the
early universe. Any deviation from 1 would be magnified by several tens of orders of 
magnitude since the Planck era, so the fact that $\Omega_0$ today is observed to be 
roughly within an order of magnitude of 1 suggests incredible fine tuning. This so-called
flatness problem is resolved by the VPP model, where $c=\dot{R}$ makes the
curvature term $\Omega_k = k$ constant, thus implying that $\Omega$, the sum
of the density parameters from all species is also constant for all time.

Moreover, since $\Omega$ is constant and $k$ is defined to take on only the 
choices 0, -1, or 1, we can see
that equation (\ref{Omega}) admits only three solutions, $\Omega$=0, 1, or 2.
As observations would appear to rule out a closed universe with 
$\Omega=2$ (and certainly an open one with $\Omega=0$), we have
the general result that $k=0$, $\nu = 4\pi /3$, and
\begin{equation}
\Omega(t) = 1.
\label{eureka}
\end{equation}

Inflation resolves the flatness problem by invoking a field with $w<-1/3$, such
as is associated with a cosmological constant. Breaking
$\Omega$ into its components and assuming constant $c$ and $G$, the SBB model yields
\begin{equation}
\frac{8\pi G}{3}\frac{\rho_{\rm m} R^2}{\dot{R}^2} + 
\frac{8\pi G}{3}\frac{\rho_{\rm r} R^2}{\dot{R}^2}
 + \frac{\lambda c^2}{3 \dot{R}^2}  = 1+ \frac{k c^2}{\dot{R}^2}
\;\;\;\;\;\;\;\;\;\;\;\; ({\rm constant}~c,~G)
\end{equation}
where all density terms now have a similar dependence on $\dot{R}$. However, SBB holds that for any
$w \geq -1/3$, the $\rho_i R^2$ terms are a decreasing function of time, meaning that
the $\Omega_\lambda$ would come to dominate over matter or radiation at late times. The fact that 
$\Omega_\lambda$ is observed to be of order unity today is thus termed the lambda 
problem. The VPP model resolves the lambda problem since $\Omega_\lambda = \lambda /3$
[c.f. equation (\ref{OmegaDefs})] is also constant for all time. Moreover, we
will see that the VPP model may not require a cosmological constant
to explain the observations of distant Type Ia supernovae.

\subsection{Conservation Laws and Scaling Relations}

From equation (\ref{G2}) we can obtain
\begin{equation}
\frac{\dot{G}}{G} = \frac{2 - 6q + v}{2-q } \frac{\dot{R}}{R}; \;\;\;\;\;\;\;\;\;\;\;\;\;
v \equiv \frac{R^2}{\dot{R}^3} \frac{d^3 R}{dt^3}
\label{Gdot}
\end{equation}
and from equation (\ref{q}) we have
\begin{equation}
\frac{dq}{dt} = - \frac{d}{dt} \left( \frac{\ddot{R}R}{\dot{R}^2} 
\right) = \frac{3}{2}\frac{dw}{dt}.
\end{equation}
Assuming for the time that $dw/dt = 0$, it follows that $v = q(2q+1)$. Plugging this result
into equation (\ref{Gdot}) gives
\begin{equation}
\frac{\dot{G}}{G} = (1-2q) \frac{\dot{R}}{R}.
\label{Gdot2}
\end{equation}
In addition, we note that 
\begin{equation}
\frac{\dot{c}}{c} = \frac{\ddot{R}}{\dot{R}}  = -q \frac{\dot{R}}{R}.
\label{dotc}
\end{equation}

If we differentiate equation (\ref{Friedmann1}) with
respect to $t$ and combine it with equation (\ref{Friedmann2}) we obtain the
conservation equation
\begin{equation}
\dot{\rho} + 3\frac{\dot{R}}{R} \left( \rho + \frac{p}{\dot{R}^2} \right)
= - \rho \frac{\dot{G}}{G} - k \frac{\dot{R}}{R} \left( \rho + \frac{3p}{\dot{R}^2} \right)
\end{equation}
which upon rearranging and using equation (\ref{Gdot2}) becomes simply 
\begin{equation}
-3 \frac{\dot{R}}{R} = \frac{\dot{\rho}}{\rho}.
\end{equation}
Alternatively, since $V = \nu R^3$, this can be written as $\rho \dot{V} = -V d \dot{\rho}$
which is precisely the equation for matter conservation, i.e.,
\begin{equation}
\frac{dM}{dt} = \frac{d}{dt}(\rho V) = 0. 
\end{equation}

This result is of course simply a restatement of Postulate 3, that the total mass of the Universe
is constant, which replaces the usual energy conservation law ($E \propto c^2$ is clearly
not conserved with varying $c$). Note that this ``mass'' however, refers
to {\em all} the components of $\Omega$ including ordinary matter, radiation, $\lambda$,
etc. The ordinary matter component is straightforward to understand. For other species $i$,
we define an ``equivalent mass'' $M_{i}$ which enters into the conservation equation.
For radiation, we have constant $M_{\rm r} = {E_{\rm r}}/{c^2}$, while for 
the cosmological constant we have constant
\begin{equation}
M_{\lambda} = \rho_\lambda V = \left( \frac{\lambda c^2}{8\pi G R^2}\right) 
\left(\frac{4\pi}{3} R^3 \right) =    \frac{\lambda}{3}M 
\end{equation}
[c. f. footnote \ref{foot1} and equation (\ref{OmegaDefs})] 
obtained after substituting equations (\ref{G2}) and
(\ref{mu}) with $w=-1$.

If we take the Universe to be comprised of a large number of systems, which
may be isolated or interacting, then two corollaries follow from Postulate 3: 
(1) the equivalent mass of any isolated system remains constant, and
(2) equivalent mass is conserved in any interaction.
If we take the example of a single non-interacting photon with equivalent mass
$m_\gamma$ as an isolated system,
we have $m_{\gamma} = E_{\gamma}/c^2 = h/ \lambda_{\gamma} c = {\rm constant}$
where $E_\gamma$ is the photon energy, 
$\lambda_{\gamma}$ is the photon wavelength and $h$ is Planck's ``constant'' which
in the VPP model is required by Postulate 3 to vary as $h(t) \propto R \dot{R}$. 
We defer the discussion of $h(t)$, however, to Section \ref{section4}. 
One result of equivalent mass conservation is that the SBB scaling laws for 
the $\rho_i$ do not 
hold, since densities for {\em all} species scale as $R^{-3}$. For example, the SBB result
that $\rho_{\rm r} \propto E/R^3 \propto R^{-4}$
is replaced by  $\rho_{\rm r} \propto E/c^2 R^3  \propto R^{-3}$.

Note that while $\Omega$ and $M$ are truly constant, 
the individual terms comprising them need not be. In other words, we have not required that
the equivalent mass associated with each particular species remains constant, only their sum.
Indeed, as physical processes convert
equivalent mass between various forms, the relative
values of the $\eta_i = \Omega_i = \rho_i / \rho = M_i / M$ will change, as will
the total pressure $p$. In the VPP
model, the effects of such processes are contained entirely in $w = \sum_i w_i \eta_i$,
which reflects a weighted average of the pressure constituents and can change over time.
Some of the results presently derived depend on the variation of $w(t)$.
In practice, small amounts 
of equivalent mass are always undergoing conversion (e.g., in stars), but such effects
are likely to be negligible, and we may take the value of $w$ in the current epoch,
denoted by $w_0$, to be effectively constant.\footnote{The present value of
$\eta_{\rm r}$, the density fraction contributed by radiation, is of order $10^{-5}$.}
In the early universe however, $w$ may have undergone more extreme
changes, for example in the transition from radiation domination to matter domination
or during earlier phase transitions associated with cosmic fields. 
In the following, we assume that $w_0$ is constant and simply note that
the theory predicts there to be distinct effects in the window
around the transition from radiation to matter domination 
(such as a slowing of the expansion speed, $c$, and a weakening
of gravity), which might potentially be measurable, for example,
in the spectrum of perturbations.

Other scaling relations change in a straightforward manner under the VPP model. 
For example, since the Friedmann equation
requires that $\rho_i R^2 /\dot{R}^2$ is constant, we have that $c = \dot{R} \propto R^{-1/2}$
and $R \propto t^{2/3}$ for all species $i$. As a result of the scaling for $c$, we 
see that energy, $E \propto c^2 \propto R^{-1}$ which is precisely the same variation 
found in the SBB model where $h$ and $c$ are assumed
constant. Thus, the familiar $R^{-1} \propto (1+z)$ redshift scaling of energy remains
precisely intact, albeit with a different rationale. Note also that in the SBB model,
the constancy of $c$ implies that frequency scales like energy as $R^{-1}$, offsetting the scaling
of wavelength, while time scales as $R^{3(1+w)/2}$ which goes like $R^2$ for radiation domination
and $R^{3/2}$ for matter domination. In the VPP model, 
frequency scales as $R^{-3/2}$, precisely the inverse scaling
of time, $t \propto R^{3/2}$. The fact that energy scales as $R^{-1}$
while frequency scales as $R^{-3/2}$ arises as a consequence of Postulate 3 and
the implied scaling behavior of Planck's
parameter, $h$, which we discuss in Section \ref{section4}.

Another consequence of the constant $R^{-3}$ scaling of the $\Omega_i$ terms is a different
result for the age of the Universe, $t_0$, which is given by
\begin{equation}
t_0 = \int_{0}^{t_0} dt = \int_{0}^{R(t_0)} \frac{dR}{\dot{R}}.
\end{equation}
Manipulation of equations (\ref{Friedmann1}) and (\ref{eureka}) yield
$\dot{R}^2 = {H_0}^2 {R_0}^3 / R$, where $H \equiv \dot{R}/{R}$ and a subscripted 0
denotes the value of a quantity at the present time $t_0$. Substituting this expression
into the equation above yields
\begin{equation}
t_0 = \frac {2}{3 H_0}
\end{equation}
independent of $w$, and of course of $\Omega$, unlike the SBB result.
Note that writing the $R(t)$ scaling relation as $R/R_0 = (t/t_0 )^{2/3}$ we have that
$c_0 = \dot{R}_0= (2/3)(R_0/t_0)$. In other words, the horizon distance, given by $d_H = R_0 = (3/2)
c_0 t_0$ is a factor of 1.5 larger than $c_0 t_0$--the distance one would obtain by simply multiplying
the present speed of light by $t_0$--due to the fact that the expansion has been slowing, i.e.,
$\ddot{R} = \dot{c} \propto -R^{-2} \propto -t^{-4/3}$. This slowing means that there can be
objects within our horizon today that are outside the sphere defined by $c_0 t_0$
and would thus appear to be receding at ``superluminal'' velocity.

One can proceed in a straightforward manner to derive other scaling relations
and cosmological quantities of interest such as the angular diameter distance,
the galaxy count-redshift relationship, etc. 
Rederiving all such quantities
is beyond the current scope, though it is hoped that other authors will do so.
The next Section will, however, focus on the derivation of
one such quantity, the luminosity distance, as it affords an interesting observational test
of the model.

\subsection{The Predicted Luminosity Distance and Implications for Type Ia Supernova Observations}

We now demonstrate that the VPP model, which predicts a
flat, $\Omega=1$ universe, appears to fit observations of distant
supernovae without needing to invoke a cosmological constant. We will begin
by deriving the expression for luminosity distance, which will differ from the 
corresponding SBB expression, and then examine corrections to other
terms entering the analysis of SN data.

Consider a source at comoving coordinate $r_1$ which emits a signal at time $t_1$
arriving at a detector at $r=0$ at time $t_0$. The RW metric in equation (\ref{RW}) gives
\begin{equation}
\int_{t_1}^{t_0} \frac{c(t)dt}{R(t)} = \int_0^{r_1} \frac{dr}{(1-kr^2)^{1/2}} \equiv f(r_1)
\label{r1}
\end{equation}
where $k=0$ from above implies $f(r_1)=r_1$.
Since $f(r_1)$ is constant, a signal emitted at $t_1 + \delta t_1$ will obey the same equation.
A simple argument then shows that
\begin{equation}
\int_{t_1}^{t_1 + \delta t_1} \frac{c(t)dt}{R(t)} = \int_{t_0}^{t_0 + \delta t_0} 
\frac{c(t)dt}{R(t)}.
\end{equation}
If we take $\delta t$ to be sufficiently small, $c(t)$ and $R(t)$ remain effectively
constant over the
negligible integration time, and we have that
\begin{equation}
\frac{\delta t_1 c(t_1)}{R(t_1)} = \frac{\delta t_0 c(t_0)}{R(t_0)} \;\;\;\;
\Longrightarrow \;\;\;\; \frac{\delta t_0}{\delta t_1} = (1+z)^{3/2}
\label{timedilation}
\end{equation}
which differs from the SBB time dilation 
result of $\delta t_0 / \delta t_1 = (1+z)$ due to the variation in $c$.
Note that unlike so-called ``tired-light'' models, which predict a time dilation factor of
unity independent of redshift, the VPP model actually predicts a {\em stronger} 
variation with redshift than the SBB model.
The result of equation (\ref{timedilation}) is critical to a correct interpretation
of cosmological data in the VPP model.
In addition to this time dependence, we have already seen that there is an energy dependence of
$(1+z)$, exactly as in the SBB model (though for different reasons), 
suggesting that a total redshift dependence for luminosity of $(1+z)^{5/2}$.

We define $L_1$ as the rest-frame luminosity of the source at $r_1$
and $F_0$ as the flux measured by the detector at $r=0$ at time $t_0$. 
The size of a sphere at time $t_0$ at the position of the detector is $4 \pi {R_0}^2 {r_1}^2 $.
Invoking Postulate 3, we define the constant equivalent mass of radiation
$M_1 = E_1/ {c_1}^2 = M_2 = E_2 / {c_2}^2$ and note that $L_1 = dE_1 / dt_1 =
2M_1 c_1 \dot{c}_1$ and $F_0 = (1/ 4 \pi {R_0}^2 {r_1}^2) dE_0 /dt_0 =
 (1/ 4 \pi {R_0}^2 {r_1}^2) 2M_0 c_0 \dot{c}_0$, so that
\begin{equation}
F_0 = \frac{L_1}{4\pi {R_0}^2 {r_1}^2} \frac{\dot{c}_0 c_0}{\dot{c}_1 c_1}
=  \frac{L_1}{4\pi {R_0}^2 {r_1}^2 (1+z)^{5/2}}
\end{equation}
where we have invoked the scaling laws previously discussed. Note that we see
the same $(1+z)^{5/2}$ dependence mentioned above.

Using the definition of $d_L$, we have
\begin{equation}
{d_L} =  \sqrt{\frac{L_1}{4\pi F_0}} = {R_0} {r_1} (1+z)^{5/4}
\label{dL}
\end{equation}
which is similar to the SBB result, 
\begin{equation}
{d_L}^{\rm SBB} = R_0 {r_1}^{SBB} (1+z)
\label{SBBdL}
\end{equation}
except for the additional $(1+z)^{1/2}$
dependence of ${d_L}^2$ arising from the time variation of $c$, and a different redshift
dependence of the $r_1$ term. In the SBB model, one must 
derive ${r_1}^{SBB} (z)$, generally via numerical methods, 
as a function of $\Omega_0$, $\Omega_\Lambda$, etc., to solve for ${d_L}^{SBB}$. In the VPP
model, one can easily solve for $r_1$ using the RW metric
\begin{equation}
r_1 = \int_0^{r_1} {dr} = \frac{1}{R(t_0)} \int_{t_1}^{t_0} c(t) dt
= \frac{R(t_0)-R(t_1)}{R(t_0)} =  \frac{z}{1+z},
\end{equation}
so that
\begin{equation}
d_L = R_0 z (1+z)^{1/4}.
\label{dL2}
\end{equation}
Note the above results for the VPP model are all exact, and independent of $w$.

The luminosity distance, however, is only one parameter entering into
the distance-magnitude relation used for measuring distant supernovae (Perlmutter \etal 1999),
\begin{equation}
{m_B}^{eff} \equiv m_R + \Delta_{corr} - K_{BR} - A_R = M_B + 5 \log{d_L} + 25
\label{md}
\end{equation}
where $d_L$ is measured in Mpc,
$m_X$ and $M_X$ denote the apparent and absolute magnitudes in waveband $X$,
$\Delta_{corr}$ is a corrective magnitude term that accounts for the observed relationship between
the width of SN light curves at a given time and their peak magnitude (enabling
their use as standard candles), $K_{XY}$ is the
cross-filter $K$-correction from the observed $Y$ band to the rest-frame $X$ band,
and $A_X$ is the extinction in the $X$ waveband, arising from dust
along the line of sight (either in our own Galaxy or in the host).
In the context of the VPP model, we must investigate the potential variation
of all terms in equation (\ref{md}). 
Rather than re-derive all additional terms, however, we consider only the differences
between equation (\ref{md}) in the SBB model vs. the VPP model.

As was the case with luminosity distance, the source of variation among terms
between the two models is the different time dilation factor in equation (\ref{timedilation}).
Taking the $A_R$ term, which can be defined as $A_R = 2.5 \log ({I_R}^{obs} / {I_R}^{em})$ where
${I_R}^{obs}$ and ${I_R}^{em}$ are the observed and emitted specific intensities, we can follow
logic similar to that used in the derivation of $d_L$ to conclude that the only difference between
$A_R$ in the two models arises from the additional $(1+z)^{1/2}$ dependence of the
time dilation factor, so that 
\begin{equation}
{A_R}^{VPP} = {A_R}^{SBB} + 2.5 \log (1+z)^{1/2}.
\label{AR}
\end{equation}
In the absence of time variation, the $K$-correction term should be identical in both models,
as it depends only on energy and wavelength, which have identical redshift dependences in the SBB
and VPP models. However, since the features of SN spectra vary with time, 
the $K$-correction term {\em does} carry an implicit time dependence 
as has been noted by Kim, Goobar \& Perlmutter (1996), Nugent, Kim, \& Perlmutter (2002),
and Davis \& Lineweaver (2003), and must therefore be recalculated.
The $K$-correction of course arises from the need to correct $R$-band 
photometry back to the rest-frame
$B$-band in order to allow to comparisons with low-redshift templates. 
To implement this correction, the data is time dilated by the usual SBB factor of $(1+z)$, 
as opposed to the VPP factor of $(1+z)^{3/2}$. Inspecting the definition
of $K_{BR}$ (Kim, Goobar \& Perlmutter 1996) we conclude that
\begin{equation} 
{K_{BR}}^{VPP} = {K_{BR}}^{SBB} + 2.5 \log (1+z)^{1/2}
\label{KBR}
\end{equation}
to reflect the additional 
$(1+z)^{1/2}$ dependence of the VPP time dilation factor. 

The remaining terms in equation (\ref{md}) do not vary between the two models.
Note that the $\Delta_{corr}$ term is actually an empirically measured term.
While it is true that the chosen parameterization for this factor, 
given by $\Delta_{corr} = \alpha (s-1)$ where
alpha is a fitted constant and the stretch factor $s$ is related to the width of the light
curve, $w$, by $s=w/(1+z)$ (see Perlmutter \etal 1999),
is motivated by the assumption
of a $(1+z)$ time dilation factor, this choice is in fact arbitrary. A different
choice of parameterization, reflecting a different time dilation factor, could have been
chosen which should in principle yield a correction of the same magnitude. 
At first glance, this would appear to be inconsistent with the results of Goldhaber \etal (2001)
who find evidence for a $(1+z)$ light-curve time-axis broadening, consistent with the SBB
model. However, these results are derived based on $K$-corrections which assume
a $(1+z)$ time dilation to begin with. A different choice of time dilation factor in $K_{BR}$,
and correspondingly different parameterization of $\Delta_{corr}$ could likely yield
a different light-curve time-axis result, as pointed out by Davis \& Lineweaver (2003).

Rewriting equation (\ref{md}) for both the SBB model
\begin{equation}
m_R + \Delta_{corr} - M_B - 25 = 5 \log{{d_L}^{SBB}} + K_{BR}^{SBB} + {A_R}^{SBB}
\label{mdSBB}
\end{equation}
and the VPP model
\begin{equation}
m_R + \Delta_{corr} - M_B -25 = 5 \log{{d_L}^{VPP}} + K_{BR}^{VPP} + {A_R}^{VPP} 
\label{mdVPP}
\end{equation}
and using equations (\ref{AR}) and (\ref{KBR}), it follows that if $5 \log({{d_L}^{VPP}}(z)) + 
5 \log (1+z)^{1/2} \approx  5 \log({{d_L}^{SBB}}(z,\Omega_0,\Omega_\Lambda))$, 
where ${{d_L}^{SBB}}(z,\Omega_0,\Omega_\Lambda)$ includes
the best-fit density parameters from observations, then the VPP model
can provide an equally good fit to the SN data.
This condition can be restated as the requirement that
${{d_L}^{SBB}}(z,\Omega_0,\Omega_\Lambda) \equiv R_0 f^{SBB}(z,\Omega_0,\Omega_\Lambda)$ is equal
to $(1+z)^{1/2} {{d_L}^{VPP}}(z) = R_0 z (1+z)^{3/4} \equiv R_0 f^{VPP}(z)$ 
where we have used equation (\ref{dL2}) and defined the function $f$ as the 
``effective distance function'' for each model. In the SBB model, this is simply
the luminosity distance divided by $R_0$, while in the VPP model an additional
$(1+z)^{1/2}$ dependence has been absorbed from the time dilation corrections.\footnote{Note the 
factor $R_0 = c_0/H_0$ in the luminosity distance is identical in both models
and therefore cancels out.} Thus, the VPP model would fit type Ia SN observations if 
$f^{VPP}(z)$ closely approximates $f^{SBB}(z,\Omega_0,\Omega_\Lambda)$ for the best-fit
values of $\Omega_0$ and $\Omega_\Lambda$.

\begin{figure}
\plotone{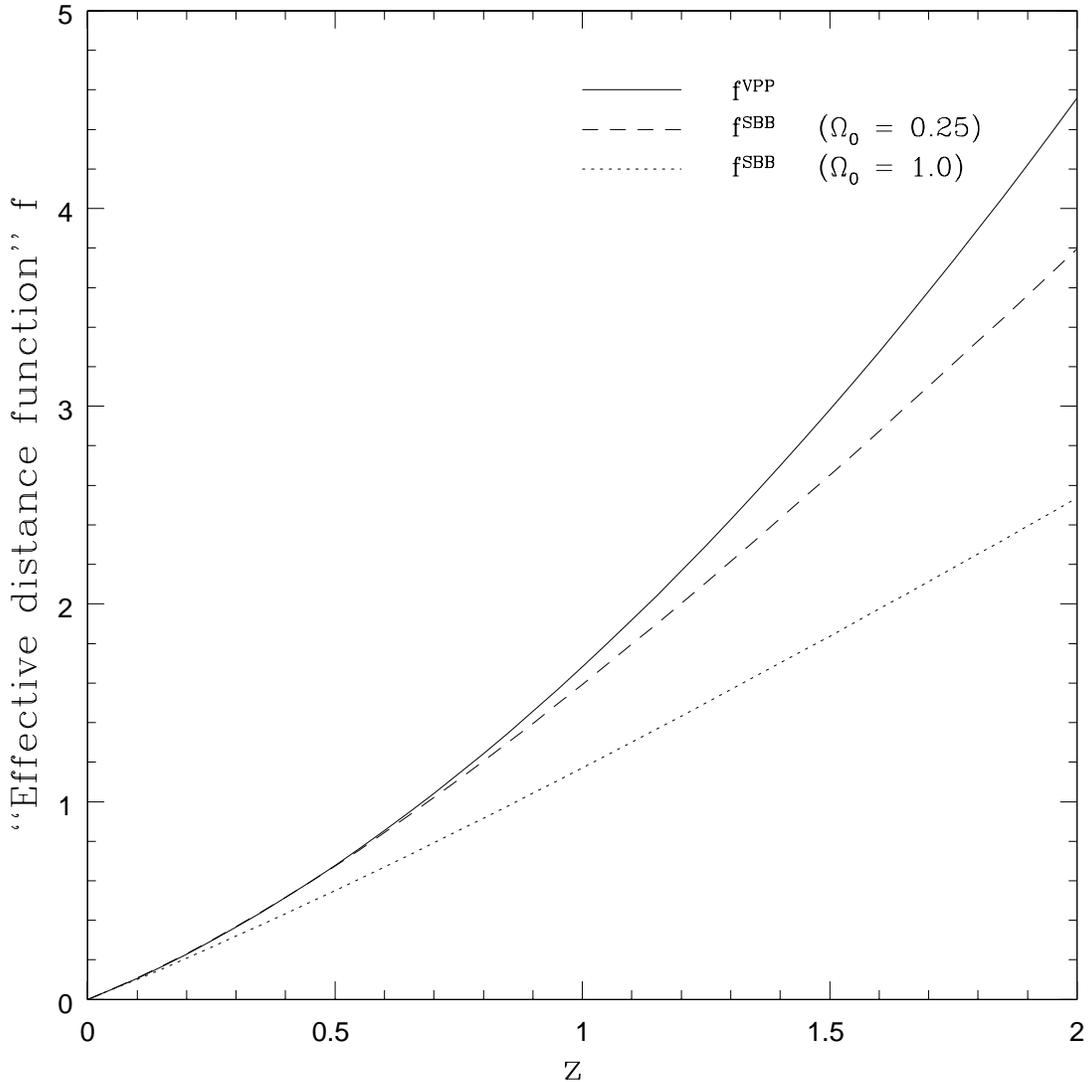}
\caption{Plot of the ``effective distance function'' $f$ defined as $f^{SBB} = {d_L}^{SBB}/R_0$ 
for SBB models and as $f^{VPP} = ({d_L}^{VPP}/R_0) (1+z)^{1/2}$ 
in the case of the VPP model to account for the impact of a different time dilation factor when
comparing with SN data (see text).
The dotted line shows $f^{SBB}$ for a flat SBB model with $\Omega_0=1$, while
the dashed line shows the $f^{SBB}$ for a flat SBB model with $\Omega_0=0.25$ and
$\Omega_\Lambda = 0.75$, the best-fit values derived from current supernova observations.
The solid line shows $f^{VPP}$, 
which is found to be within 2\% of the best-fit SBB result from $0 < z < 0.7$ and within
10\% from $0.7 < z < 1.3$, indicating it provides a comparable fit to the SN data.}
\label{fig:SN}
\end{figure}

Figure \ref{fig:SN} displays $f^{VPP}(z)$, represented by the solid line, as well as 
$f^{SBB}(z)$ for two choices of cosmology: (1) a flat,
$\Omega_0 =1$ model represented by the dotted line, and (2) a flat model with
$\Omega_0 = 0.25$ and $\Omega_\Lambda = 0.75$, in line with the best-fit SBB values
from observations (Tonry 2003; Knop 2003), represented by the dashed line.
The values for $f^{SBB}(z)$ were obtained from an analytic expression derived
by Pen (1999). The best-fit $f^{SBB}(z)$ with $\Omega_0 = 0.25$ shows that the SBB
luminosity distance measured from SN observations is roughly
20\% greater than that for a flat $\Omega_0=1$ universe at $z \approx 0.3$
and about 40\% greater at $z \approx 1$, and thus requires invoking a cosmological
constant to produce acceleration.

Remarkably, $f^{VPP}(z)$ is within 2\% of the best fit $f^{SBB}(z)$ 
out to $z = 0.7$ and within 10\% out to $z= 1.3$, and therefore seems to do a comparable
job of explaining the data! We reiterate that the VPP model
requires $\Omega = 1$ and nowhere have we required that a cosmological constant
be contributing to $\Omega$. 
If the assumed impact of the VPP time dilation factor on the various terms 
in equation (\ref{md}), specifically $K_{BR}$, is correct, then it would appear
the VPP model can explain type Ia SN observations in a flat, matter-dominated
universe without invoking a cosmological constant.
It is interesting to note that under the present assumptions 
$f^{VPP}(z)$ grows faster than the best-fit $f^{SBB}(z)$ as redshift increases, 
with a 20\% greater
value at $z=2$. This would predict that as higher-$z$ supernova measurements
are made available, higher values of $f^{SBB}$ will be observed that would require
even larger values of $\Omega_\Lambda$ under the SBB model.

Other interesting predictions might be derived from an application of the present model
to the thermal history of the early Universe, nucleosynthesis and element abundances, the
cosmic microwave background radiation,
and the theory of large-scale structure formation.\footnote{For example, 
since the VPP model predicts effectively the present-day value for $G$, but a higher value for $c$,
during the era of structure formation, it may provide a way to suppress power on small scales since
the radiation feedback on these scales would have higher energy and pressure than assumed in the
SBB model.} While such investigation is 
outside the present
scope, it is hoped that other authors might address these and other areas of interest.

\section{Quantitative Behavior of $G(t)$ and $c(t)$} \label{section3}

We now turn to a quantitative examination of our results for $G(t)$ and $c(t)$, 
in order to understand whether the predicted
variation of the physical parameters in our model might be confronted by
observations. We will see that among the consequences of the connection of the physical
parameters with the cosmic expansion will be that our Universe satisfies the Schwarzschild
equation.

We begin by examining our postulate that $c=\dot{R}$. Using the current best value for $H_0$
of 71 km s$^{-1}$ Mpc${^-1}$, we can calculate $R_0 = c_0 /H_0$ to obtain the familiar result
of $1.3 \times 10^{28}$ cm. Equation (\ref{dotc}) yields
\begin{equation}
\dot{c_0} = - \frac{q {c_0}^2}{R_0} = -6.9 \times 10^{-8}q~~{\rm cm~s}^{-2}.
\end{equation} 
Recall that $q = (1/2)(1+3w)$ which is $1/2$ in the case of matter domination. 
Note that $\dot{c}=0$ (uniform expansion) for $w=-1/3$ 
and is greater than 0 for any value below this. A cosmological
constant, with $w=-1$, would thus produce acceleration as expected, though as we have seen,
nothing in the model suggests the need for a nonzero $\lambda$. Density constituents
with $w > -1/3$ result in a decelerating universe.
For constant $q$ (i.e., $dw/dt=0$), higher-order derivatives of $c$ are given by
\begin{equation}
\frac{d^n c}{dt^n} = \left( \prod_{i=1}^n \left[ n(q+1) - 1 \right] \right) (-1)^n 
\frac{c^{n+1}}{R^n}.
\end{equation}

Rewriting equation (\ref{G2}) using all of our results to this point gives
\begin{equation} 
G(t) =  \frac{\mu \nu}{M} (6Rc^2 + 3R^2 \dot{c}) =  \frac{1}{2M (2-q)} (2Rc^2 + R^2 \dot{c})
= \frac{Rc^2}{2M}
\label{G3}
\end{equation}
which is precisely the Schwarzschild equation, with $R$ and $M$ characterizing the scale
factor and mass of the Universe! Note that this result is independent of the constituents
of $\Omega$ and also independent of $w$. Conversely, it can be shown that substitution
of Postulates 1 and 2, using equations (\ref{G2}) and (\ref{mu}) for $G$, 
into the Schwarzschild equation will in turn
produce the Friedmann equation:
\begin{equation}
R_{\rm Sch} = \frac{2GM}{{c}^2} = \frac{2 \mu \nu }{\dot{R}^2}
(6R\dot{R}^2 + 3R^2 \ddot{R}) = R(t)
\end{equation}
which can be rewritten as 
\begin{equation}
2 \mu \nu \left(6 + \frac{3{R}\ddot{R}}{\dot{R}^2} \right) = 1,
\end{equation}
and is thus precisely equal to the transformed Friedmann equation (\ref{modFriedmann})
with $\nu = 4\pi/3$ for a flat, $\Omega=1$ universe.
Various authors have also previously pointed out that the 
physical radius, $R_0= c_0/H_0$, of a flat $\Omega_0 = 1$ universe 
is precisely equal to the Schwarzschild radius, 
\begin{equation}
R_{\rm Sch} = \frac{2GM}{{c_0}^2} = \frac{2G}{{c_0}^2} \frac{4\pi R_{0}^3}{3} 
\Omega_0 {\rho_c}_0 = \frac{ {R_0}^3 {H_0}^2 }{{c_0}^2} = R_0
\label{Rsch}
\end{equation}
where ${\rho_c}_0 = 3{H_0}^2/8{\pi} G$ is the current value of the critical density.
This correspondence between an Einstein-de Sitter
universe and a black hole is supported by the VPP model, which demonstrates that 
the Friedmann and Schwarzschild equations specify the same dynamics, and
will be further elaborated upon in the
discussion of Planck's parameter in Section \ref{section4}.
Note that using 
the known value of $G$ allows us to calculate $M = c^2 R/2G = 8.8 \times 10^{55}$ g.

Returning to equation (\ref{G3}), we find that 
there are only 2 cases in which $G=0$, namely, the solution with $R=0$ or 
a static universe with $c=0$. 
The degree of variation in $G$ can be expressed using 
equation (\ref{Gdot2}) to obtain 
\begin{equation}
\dot{G} = (1-2q)G \frac{\dot{R}}{R} = 1.53 \times 10^{-25} (1-2q)~~{\rm g}^{-1}~{\rm
cm}^{3}~{\rm s}^{-3}.
\end{equation}
For a matter-dominated Universe, $1-2q = 0$, so that $\dot{G}$ vanishes and $G$ resumes the status
of a true constant. Note however, that this would not be true for all time since, for example,
$1-2q = -1$ during the prior epoch of radiation domination. Interestingly, this suggests
that $G$ was larger in this earlier epoch and
decreased to roughly its present value 
as $\eta_{\rm r}$ became small after matter-radiation equality
(where we are assuming no other significant species besides matter and radiation). 
Other than the case of matter domination, $\dot{G}$ is always nonzero except in the trivial 
solutions noted above for $\dot{c}$. For $dw/dt=0$,
higher-order derivatives of $G$ are given by
\begin{equation}
\frac{d^n G}{dt^n} = \left( \prod_{i=1}^n \left[ -(n+1)q - (n-2) \right] \right)  
\frac{G c^n}{R^n}.
\end{equation}

\section{The Planck Parameter $h(t)$} \label{section4}

We have already noted that application of Postulates 1 and 3 to radiation
implies that $h \propto M R\dot{R}$. In order to determine the constant
of proportionality, we note that a unit of length--the so-called
Planck length, $d_{\rm Pl}$, taken to be $d_{\rm Pl} = \sqrt{Gh/c^3}$--can 
be constructed from a combination of $h$, $G$, and $c$, and therefore that
\begin{equation}
h = \frac{ {d_{\rm Pl}^2}}{R^2} \frac{R^2 \dot{R}^3}{G}. 
\label{h}
\end{equation}
We define $N =  R^2 /{d_{\rm Pl}^2}$ 
and note that $R \dot{R} = \dot{A}/ 2 \sigma$, 
where $A$ is surface area of the Universe and $\sigma$ is the geometric
factor defined by $A=\sigma R^2$ and is equal to $4\pi$ in the case
of the VPP model which requires a flat, $\Omega=1$ universe.
Substituting equation (\ref{G3}) for $G$ into (\ref{h}), we obtain
the corollary defintion for $h$:

\begin{equation}
h = \frac{2MR\dot{R}}{N} = \frac{M\dot{A}}{4\pi N}.
\label{P5}
\end{equation}

Like $G$, $h$ vanishes only for $R=0$ or the static $c=0$ case.
The time variation of $h$ is given by
\begin{equation}
\frac{\dot{h}}{h} = (1-q)\frac{c}{R} 
\end{equation}
so that the present variation is 
\begin{equation}
\dot{h_0} = 1.53 \times 10^{-44} (1-q)~~{\rm g~cm}^{2}~{\rm s}^{-2}. 
\end{equation}
Note that $\dot{h}$ vanishes for a radiation-dominated universe and is positive
for a matter-dominated universe, suggesting that $h$ is increasing today. 
For $dw/dt=0$, higher-order derivatives of $h$ are given by
\begin{equation}
\frac{d^n h}{dt^n} = \left( \prod_{i=1}^n \left[ (1-q) -(n-1)(1+q) 
\right] \right) \frac {h c^n}{R^n}.
\end{equation}

We have already seen that the VPP model describes the Universe as a black hole.
Bekenstein was the first to draw the link between entropy, $S$, which 
characterizes the number of degrees of freedom of a system, and the area $A$ of a black hole.
This relationship was later made explicit by Hawking (1976).
Using the classical definition of entropy as an analogy, we define a new quantity,
$C$, termed ``cosmic entropy''  given by
\begin{equation}
C = h N = \frac{M \dot{A}}{4\pi}
\label{cosmic entropy}
\end{equation}
where $N=  R^2 /{d_{\rm Pl}^2} = 1.04 \times 10^{121}$ is the 
(fixed) number of degrees of freedom--equal to 
the number of Planck area units 
on the surface of the Universe--and $h$ is the (evolving) unit value of cosmic entropy, 
or equivalently, the mass per degree of freedom,
times the time rate of change of area.

From Postulate 3 we have
\begin{equation}
\dot{\rho}V + \rho \dot{V} = 0
\end{equation}
which upon differentiating $\rho$ and
multiplying by $wc^2$ can be rewritten as the familiar thermodynamic
relation
\begin{equation}
pV = wE
\label{thermo1}
\end{equation}
which of course yields the familiar results $p=0$ for matter and $p= E/3$ for radiation.
Taking the time derivative of equation (\ref{cosmic entropy}) we have
\begin{equation}
\frac{dC}{dt} = \frac{M}{4\pi} \left[ 2 \sigma ({\dot{R}^2} + R\ddot{R})\right] = 2Mc^2 (1-q) =  
2E (1-q)
\label{thermo2}
\end{equation}
where $Mc^2 = E$ is the total energy of the Universe. Note that since $Mc^2$ is positive definite,
$dC/dt$ is positive for all $q<1$, which corresponds to $w < 1/3$, and vanishes for
$w=1/3$, as is the case for radiation. 
Since we know of no species with $w>1/3$, we may conclude that cosmic entropy can
never decrease.

It is interesting to note that in the VPP model, $c$--the parameter central to special
relativity--is associated with a change in length, $h$--the parameter central to quantum 
mechanics--is associated with a change in area, and $G$--the parameter central
to general relativity--is associated with a change in volume.
The holographic principle put forth by t'Hooft (1993) states that the information contained
in a volume of space can be represented by a theory specified entirely on its surface and
that there is at most one degree of freedom per Planck area. It appears the VPP model might
support this principle. Consider the apparent symmetry between the following two
statements, suggested by equations (\ref{thermo1}) and (\ref{thermo2}): 
(1) radiation does not generate a change in cosmic entropy, but does generate pressure which drives a
change in $G$, a parameter proportional to the $N-1$ time derivative of an 
$N$ dimensional space ($N=3$), and central to a theory (general relativity)
most closely associated with the behavior of matter.
(2) matter does not generate a change in gravity, but does 
generate cosmic entropy which drives a
change in $h$, a parameter proportional to the $N-1$ time derivative of an 
$N$ dimensional space ($N=2$), and central to a theory 
(quantum mechanics)
most closely associated with the behavior of radiation.
This would appear to suggest an equivalence between the ``3D'' theory of general
relativity and the ``2D'' theory of quantum mechanics.
Of course, both general relativity and quantum mechanics can be applied to the
description of both matter and radiation, but there remain inconsistencies between the two
whose resolutions are of course the
subject of string theory. While it is beyond the present scope to investigate the string
theoretic implications of the VPP model, it is hoped that others might 
elucidate these.

It is interesting to interpret Postulate 3 in light of the thermodynamic behavior of black holes.
Hawking's Area Theorem states that, classically, the area (determined uniquely
by the mass) of a black
hole never decreases.\footnote{Since $\dot{A} = 8\pi R c \geq0$, this in turn implies that the
Universe never contracts, consistent with our result that $k=0$.}
Quantum mechanically, Hawking showed that black holes do emit blackbody
radiation with a temperature proportional to the inverse of their mass, 
with roughly 1 photon
emitted per crossing time. For a black hole with the mass of the Universe, the temperature
is of course infinitesimally small and the decay lifetime immeasurably large.
This would suggest that, effectively, $M$ does not decrease. If our Universe is an isolated system,
it stands to reason that $M$ would also never increase, and thus that
$dM/dt = 0$, as in Postulate 3. Some string theoretic models propose the existence of
multiple universes in so-called bubble-universe models. In this context, it may be possible that the
condition $dM/dt=0$ holds for a period of time when our Universe is isolated, but
may be violated when our Universe comes into contact with another, which would obviously
have strong implications for the current model.

\section{Conclusion}

This work has put forth several radical postulates on the definitions and variations
of physical parameters, and drawn a number of conclusions which are different
from those of standard physics. As mentioned in the Introduction, 
it has not presented a complete theoretical framework in which these ideas can be fully
embedded. However, it is hoped that a viable path for further inquiry
has been laid out for both theoretical and experimental investigation.
In this concluding section we highlight a few starting points for such
an inquiry. 

From a theoretical perspective, there is nothing which prevents the fundamental
``constants'' from evolving. The VPP model is based on the Machian
premise that physical parameters reflect global characteristics of the Universe,
and are therefore {\em expected} to evolve in accordance with the expansion.
In practice, there are of course observational constraints on the variation of these
quantities. To the degree that ``standard'' physics has arisen from laboratory experimentation
conducted over a cosmologically infinitesimal timeframe, it can be viewed simply
as the specialized case where $G(t) \longrightarrow G_0$, $c(t) \longrightarrow c_0$,
$h(t) \longrightarrow h_0$, etc., provided that $G(t)$, $c(t)$, and $h(t)$ do not violate
experimental constraints. The present model is only one in a long history of models that
have been proposed under this rationale.

At first glance, the VPP model might appear to be at odds with special or general 
relativity, but in fact this may not be the case.
Consider special relativity (SR) which is based on the concept of
inertial reference frames and follows from the postulates that: (1) the laws of physics
are the same in all inertial reference frames, and (2) the speed of light as measured by any
observer in an inertial frame is constant at $c$. The VPP model, however, posits the existence of
a preferred frame in which we and everything else exists, so how can these be reconciled? 
Consider a classical particle with arbitrarily small mass moving at a speed $v$ that is arbitrarily
close to $c_0$. 
The VPP model states that since this particle is moving at $c_0$ in the preferred frame,
an observer in an ``inertial'' SR frame, will also measure a speed of $c_0$. Apparently,
what SR calls an inertial frame is in the VPP model any frame at time $t_*$ 
in which the speed of a massless particle is measured to be $c(t_*)$.
Under this equivalence, 
the first postulate of SR still holds and the second postulate is then merely a 
direct consequence of the first. 
SR, however, also requires local and global energy conservation, whereas the VPP
model violates energy conservation stating that $E(t) \propto c(t)^2 \propto 1/R(t)$.
Quantifying this violation, we find that for a system of {\em any} equivalent mass
\begin{equation}
\frac{\dot{E}}{E} = 2\frac{\dot{c}}{c} = -2q\frac{c}{R}
\end{equation}
which implies a present-day percent variation in $E$ of order $10^{-18}$ s$^{-1}$,
or nearly one part in a billion per decade, using the values 
of $c_0$ and $R_0$. Thus, the predicted degree
of non-conservation may not be at odds with the predictions
of SR in terrestrial laboratory experiments.

The VPP model is of course assumed to be consistent with General Relativity (GR) by virtue
of Postulate 4. It is hoped that a suitable generalization of GR can prove this. GR
of course treats all reference frames on equal footing, but only defines reference frames
in a local sense, and thus does not have a global energy conservation law. The results
of the SBB model are derived by assuming a global energy conservation law for the Universe,
in combination with the Einstein equations. The VPP model instead assumes, via Postulate 3,
a global mass conservation law, which has been shown in Section \ref{section2}
to be consistent with the
assumed forms of the physical parameters. It is noted that in principle it might be possible to 
formulate an alternative, but equivalent, VPP model in which energy is conserved and $M=M(t)$.
If this is possible, however, such a model would seem to lack the same physical intuition
and would not appear to imbue the physical parameters with the same geometric significance
as in the current model. 

As a final comment, it is noted that the statement $c = \dot{R}$ suggests the possibility
that a generalized theory consistent with the results of the VPP model might
somehow couple massless particles directly to the cosmic expansion, so that they are
in effect swept along geodesics by the expansion. By contrast, massive particles would be
only weakly coupled to the expansion--governed on small scales by local
forces that dominate over the expansion, and only at large scales by the expansion dynamics.
 
From a theoretical standpoint, the VPP model appears attractive in that it predicts
$\Omega=1$ for all time, explains a range of cosmological
problems without invoking inflation, appears to fit Type Ia supernovae observations
without invoking a cosmological constant, and gives rise to a series of other results
that suggest the possibility of a deeper theoretical significance.
Experimentally, the VPP model appears to offer several potentially testable predictions, 
some of which have been described in this paper. 
Whether this model proves to be correct, incomplete, or wrong, it is hoped
this work will stimulate further discussion 
that may lead towards a greater understanding.

\bigskip
The author would like to thank E. Blackman for helpful discussions and comments.



\end{document}